\documentclass[11pt, a4paper]{article}
\usepackage{authblk}
\usepackage{enumitem}
\usepackage{graphicx}
\usepackage{dcolumn}
\usepackage{bm}
\usepackage{setspace}
\usepackage{amsfonts}
\usepackage[normalem]{ulem}
\usepackage{todonotes}
\usepackage[font={small}]{caption}
\usepackage[margin=2.5cm]{geometry}
\usepackage[superscript]{cite}

\begin{document}

\newcommand{\dmdt}{\ensuremath{\mathrm{d}m/\mathrm{d}t}}
\newcommand{\dx}{\mathrm{d}x}
\newcommand{\dy}{\mathrm{d}y}
\newcommand{\nife}{Ni$_{80}$Fe$_{20}$}
\newcommand{\lmesh}{l_{\mathrm{mesh}}}
\newcommand{\fres}{$f_{\mathrm{res}}$}
\newcommand{\hext}{H_{\mathrm{ext}}}
\newcommand{\figo}{\textbf{Figure}}
\newcommand{\fig}{Figure}
\newcommand{\hp}{H_{\mathrm{P}}}

\setcounter{Maxaffil}{4}

\title{\Large \bf Resonance-based Detection of Magnetic Nanoparticles and Microbeads Using Nanopatterned Ferromagnets}

\author[1,*]{\rm Manu Sushruth}
\author[2]{\rm Junjia Ding}
\author[3]{\rm Jeremy Duczynski}
\author[1]{\rm Robert C.~Woodward}
\author[1]{\rm Ryan A.~Begley}
\author[4]{\rm Hans Fangohr}
\author[3]{\rm Rebecca O.~Fuller}
\author[2]{\rm Adekunle O.~Adeyeye}
\author[1]{\rm Mikhail Kostylev}
\author[1,**]{\rm Peter J.~Metaxas}
\affil[1]{School of Physics, M013, University of Western Australia, 35 Stirling Hwy, Crawley WA 6009, Australia.}
\affil[2]{Information Storage Materials Laboratory, Department of Electrical and Computer Engineering, National University of Singapore, Singapore-117576, Singapore}
\affil[3]{School of Chemistry and Biochemistry, M310, University of Western Australia, 35 Stirling Hwy, Crawley WA 6009, Australia.}
\affil[4]{Engineering and the Environment, University of Southampton, Southampton, SO17 1BJ, United Kingdom}
\affil[*]{\textit {manu.sushruth@research.uwa.edu.au}}
\affil[**]{\textit {peter.metaxas@uwa.edu.au}}

\renewcommand\Authands{ and }

\date{\today}

\maketitle


\begin{abstract}
Biosensing with ferromagnet-based magnetoresistive devices has been dominated by electrical detection of particle-induced changes to the devices' static magnetic configuration. There are however potential advantages to be gained from using field dependent, high frequency magnetization dynamics for magnetic particle detection. Here we demonstrate the use of nano-confined  ferromagnetic resonances in periodically patterned magnetic films for the detection of adsorbed magnetic particles  with diameters ranging from 6 nm to 4 $\mu$m. The nanopatterned films contain arrays of holes which can act as preferential adsorption sites for small particles. Hole-localized particles act in unison to shift the resonant frequencies of the various modes of the patterned layer with shift polarities determined by the localization of each mode within the nanopattern's repeating unit cell. The same polarity shifts are observed for  a large range of coverages, even when hole-localized particles are covered by quasi-continuous particle sheets. For large particles however, preferential adsorption no longer occurs, leading to resonance shifts with polarities which are independent of the mode localization. Analogous shifts are seen in continuous layers where, for small particles, the shift of the layer's fundamental mode is typically about 10 times less than in patterned systems and induced by  relatively weak fields emanating  beyond the particle in the direction of the static applied field. This highlights the importance of having confined modes consistently positioned with respect to nearby particles.
\end{abstract}

\maketitle

\section{Introduction}

\begin{figure}[htbp]
\centering
	\includegraphics[width=8cm]{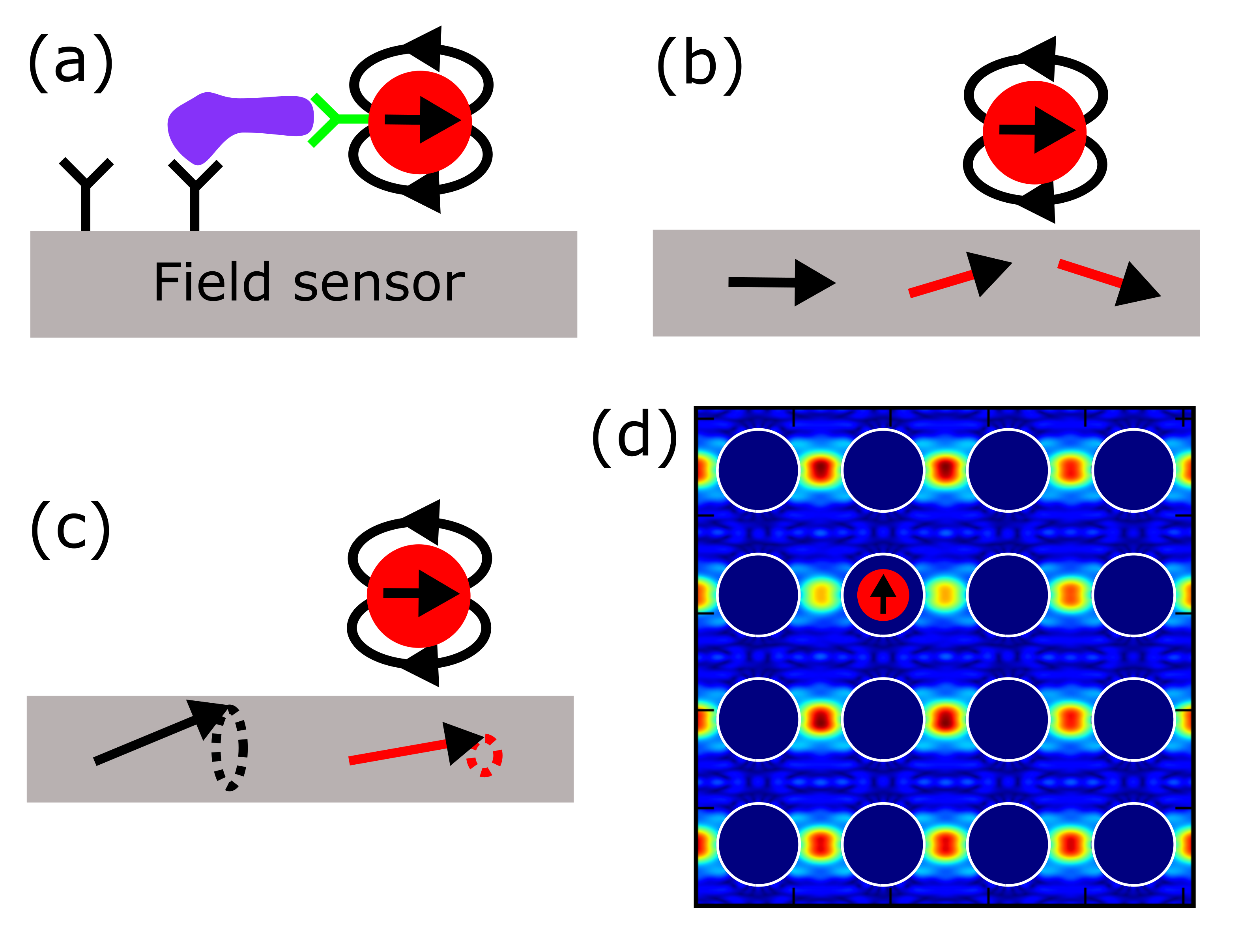}
	\caption{(a) Schematic of a sandwich immunoassay showing a magnetically labeled biological entity specifically bound to a chemically functionalized sensor. The nanoparticle stray field can induce changes to the (b) static magnetization configuration of a ferromagnetic sensing element or (c) its precessional magnetization dynamics. (d) Shows a magnetic particle (red) within a hole (white circles) in a hole-based, periodic magnonic crystal with geometry equivalent to that studied in Sec.~2.2 (300 nm wide holes). Both the particle and patterned film are magnetized towards the top of the page. The particle stray field locally modifies the amplitude of  precessional magnetization dynamics around it. The dynamics correspond to a side mode resonance which is localized between the holes. The color encodes the amplitude of the magnetization precession in the MC (red = high; blue = low). Except for the single nanoparticle, there is no magnetic material within the holes. }
	\label{f1}
\end{figure}

Magnetic biosensing techniques have shown great promise in terms of providing a matrix-insensitive biosensing platform for future use in point-of-care medical diagnostics\cite{Baselt1998,Gaster2009}. These techniques center on the detection of magnetic particles which are used as tags for analytes of interest in biological fluids (Fig.~\ref{f1}(a)). While numerous methods for detection of nanoparticles exist\cite{Chemla2000,Besse2002,Miller2002,Ejsing2005,Donolato2009,DiMichele2011,Hira2012,Devkota2013,Chung2013,Nikitin2007}, magnetoresistive structures, exploiting giant or tunneling magnetoresistance phenomena have garnered significant interest\cite{Llandro2007,Srinivasan2009,Osterfeld2008,Hall2010,Srinivasan2011,Freitas2012}. Conventional sensing with such structures is achieved via the detection of magnetic-particle-induced changes to the static magnetic configuration within these (typically multilayered) ferromagnetic devices (Fig.~\ref{f1}(b)), a process which exploits their magnetic-field dependent resistivity\cite{Lee2016}. Thus, like Hall effect sensors\cite{DiMichele2011,Hira2012}, they offer a sensing method based on detecting changes to a d.c. voltage level.

Recently, attention has begun to turn towards the possibility to exploit the magnetic field dependence\cite{Mizushima2010,Inoue2011} of resonant magnetization dynamics for particle sensing\cite{Ryan2011,Braganca2010,Fried2016, Metaxas2015}. The consequence of this dependence is that the precession frequency of the magnetization, typically in the giga-Hertz range, will be altered by particle-generated magnetic fields (Fig.~\ref{f1}(c)), potentially enabling an intrinsically frequency-based rather than amplitude-based sensing technique \cite{Braganca2010}. Notably, the stray fields generated by the particles act directly on the resonance, not requiring particle-induced changes to the magnetic ground state of the system for detection. Indeed, there are number of reasons to motivate the investigation of such methods. For example, dynamics can be induced and measured even at large external fields \cite{Petrie2014} where particle moments, and thus particle-generated fields (the basis of particle detection) can be maximized. Furthermore, using (e.g.) spin torque oscillators,\cite{Braganca2010} it is possible to efficiently  probe and read out such dynamics electrically in real time\cite{Krivorotov2005,Suto2010,Zeng2012a,Grimaldi2014} (potentially enabling high speed sensing for applications such as cytometry\cite{Loureiro2009,Loureiro2011,Helou2013}). Furthermore, in the latter devices, favorable signal to noise ratios can be retained down to sub-100 nm device length scales\cite{Braganca2010}. Indeed here we will demonstrate particle detection using a particular resonance confined laterally to a region with size on the order of $100 \times 100$ nm$^2$.

Magnonic crystals (MCs) are attractive systems to explore the fundamentals of such a sensing technique. MCs are magnetic materials which have been artificially patterned (typically at the nanoscale) to  control spin-wave (magnon) propagation or enable periodic confinement of  nano-localized ferromagnetic resonances\cite{Krawczyk2014} (e.g.~Fig.~\ref{f1}(d)). High levels of periodicity over large length scales mean that these localized resonant modes (analogous to those excited in isolated nanostructures) can be probed easily in laboratory settings in macroscopic samples. Beyond their envisioned applications in data and signal processing\cite{Lenk2011}, we have previously demonstrated that MCs can be used to understand the influence of magnetic fields generated by magnetic nanoparticles  on highly localized (or `confined') ferromagnetic resonances.\cite{Metaxas2015} Notably, using a hole-based structure enables the localization of particles within the holes with a subsequent predictable local modification of the resonant precession of the magnetization in the neighborhood of the captured particle. An example of this can be seen in Fig.~\ref{f1}(d) as a change in the color-coded precession amplitude to the left and right of the hole-localized nanoparticle.  More recently, similar effects were confirmed numerically in another hole-based MC geometry\cite{Manzin2016}.

In this paper we demonstrate the successful use of nano-localized ferromagnetic resonances for the detection of magnetic particles with a range of diameters spanning 3 orders of magnitude: from sub-10 nm superparamagnetic iron-oxide nanoparticles to 4 $\mu$m wide  magnetic beads. This particle size range also approximates a correspondingly large range of biological length scales, from single proteins to  cells. We will explicitly show the advantage of using hole based structures for the localization of both modes and particles, the latter ensuring a common, and thus reinforcing effect, from individual particles. Indeed we show that the sensitivity is an order of magnitude less when using an unpatterned continuous ferromagnetic layer. There, the dominant effect is from the  stray field extending far outside the particles which is weak compared to that  beneath or directly neighboring each particle. When particles are within the MC's holes the localization of different modes around each hole in the patterned case defines the sign of the frequency shift. This shift polarity is  maintained and its amplitude increased at high coverages where quasi-continuous particle sheets are formed. However by increasing the particle size to a degree in which the particle cannot enter the holes, we lose the mode-dependent shift polarities with all modes behaving similar to the fundamental mode of a continuous layer in that their frequencies all increase. Varying the particle size can thus enable a transition to a film-like behavior, albeit with multiple modes exisiting in the MC. 
Note that all particles studied in this work exhibit a quasi-null magnetic moment in zero field (as checked via magnetometry) and thus minimal agglomeration.

\section{Results and discussion}

\subsection{Continuous layers} 

\begin{figure*}[htbp]
\centering
	\includegraphics[width=16cm]{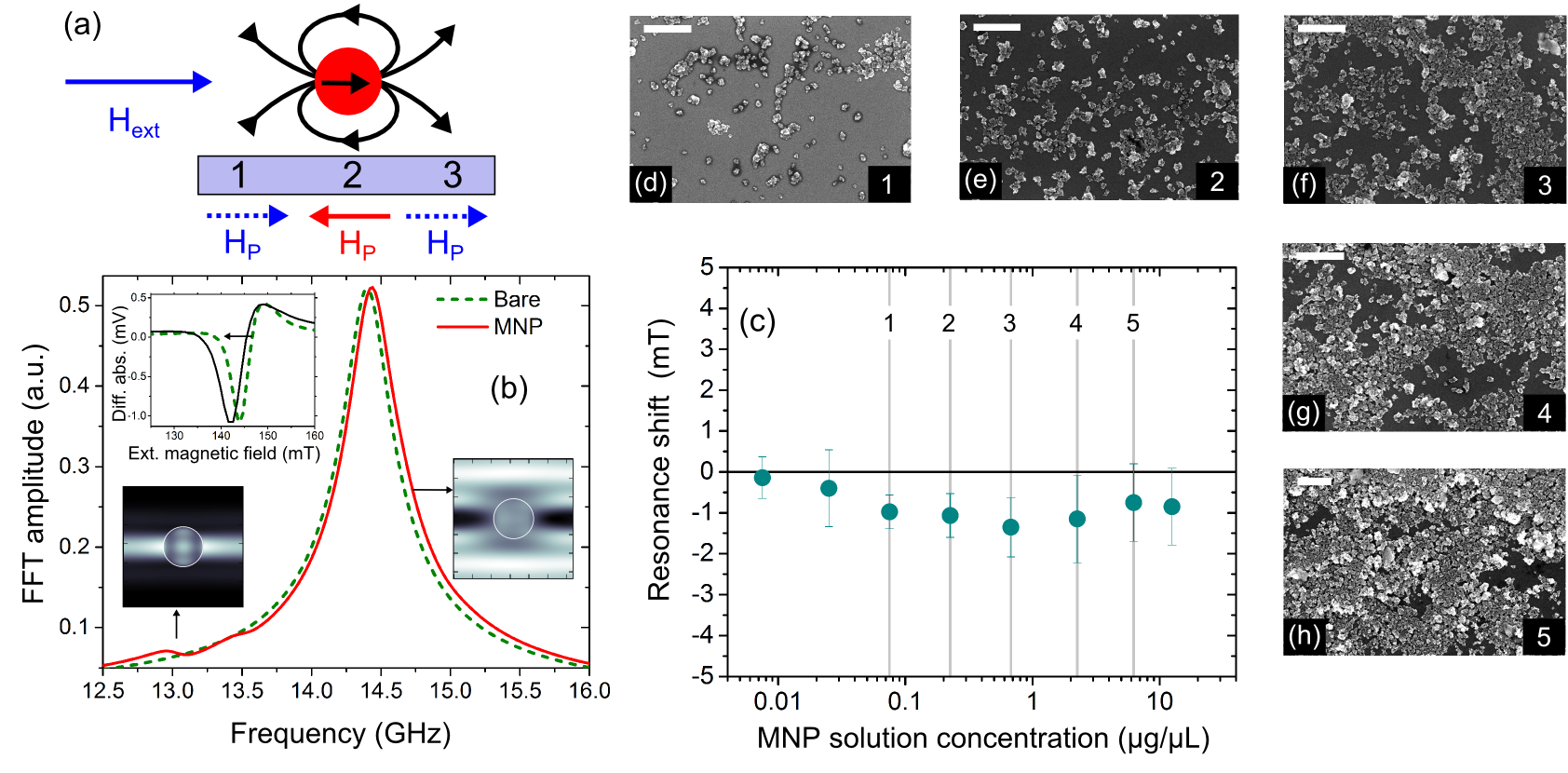}
	\caption{(a) Schematic side view of a particle placed on an underlying magnetic film. The in-plane stray magnetic field generated by a particle, $\hp$, is strong and opposes the applied in-plane static field ($\hext$) directly below it, resulting in a strong downward shift in the resonance frequency in that region (marked as `2'). In regions `1' and `3', $\hp$ is weaker (blue dashed lines) but reinforces $\hext$, thus slightly increasing the resonance frequency.  (b) Fourier transformed time domain simulation data for an un-patterned film showing a small upward shift in fundamental mode's frequency in presence of a particle. The upper-left inset shows experimental FMR traces, obtained at 12 GHz, with (solid red line) and without (dotted green line) particles (concentration is 0.675 $\mu$g.$\mu$L$^{-1}$)  on the surface of unpatterned film. Lower insets show the simulated mode profiles of the two main modes that exist in the presence of a particle. (c) Experimentally obtained result for fundamental mode shifts of an unpatterned film as a function of particle concentrations. The measured shifts for each concentration are the average shift for frequencies ranging 11.5 - 16 GHz. (d-e) SEM images showing the distribution of particles on the unpatterned film's upper surface. White scale bars are 1 $\mu$m long.} 
	\label{funiform}
\end{figure*}

We first consider the case of particle-induced resonance shifts in continuous, \textit{unpatterned} magnetic layers where resonant dynamics are typically laterally uniform across the layer.
Fig.~\ref{funiform}(a) shows a schematic of the dipole field of a magnetic particle acting on an underlying magnetic film subject to an external in-plane static field, $\hext$. The in-plane component of the particle's field, $\hp$, is strong and opposes $\hext$ directly below the particle (in region `2',   Fig.~\ref{funiform}(a), and to the sides of the particle, the latter due to the  symmetry of the dipolar $\hp$). Elsewhere (regions `1' and `3', Fig.~\ref{funiform}(a)) the in-plane component of $\hp$ is weaker but reinforces $\hext$. Subsequently, although a bare, thin magnetic layer has one primary resonance mode (which corresponds to the spatially uniform fundamental mode, shown  as a solid line in Fig.~\ref{funiform}(b)), an extra, low frequency mode appears in simulation when isolated particles are on top of the film. Note that these simulation results assume one 150 nm particle 10 nm above every $450 \times 450$ nm$^2$ of film (further simulation details are given in experimental section). 

The low frequency mode is located beneath the particle and to its sides where $\hp$ opposes $\hext$ (see bottom left inset in Fig.~\ref{funiform}(b) while comparing to region `2' in Fig.~\ref{funiform}(a)). Since the total field is reduced at that position, so is the resonance frequency. The majority of the dynamics within the layer are however concentrated in the remainder of the computation region as seen in the bottom right inset in Fig.~\ref{funiform}(b) (these areas correspond to regions `1' and `3' in Fig.~\ref{funiform}(a)). This mode is  slightly upshifted due to the weak $\hp$ in those regions reinforcing $\hext$, thus increasing the total field. Notably, at a particle-layer separation of 10 nm,  the maximum in-plane $\hp$ directly below a single, perfectly spherical particle will be approximately 15 times higher than the maximum in-plane $\hp$ at the upper/lower boundaries of the computation cell. This explains the disparity in the shift amplitudes seen for the dynamics concentrated in region `2' versus those concentrated in regions `1' and `3'.

In a field-resolved experimental FMR trace (obtained at a fixed frequency, upper-left inset of Fig.~\ref{funiform}(b)), the particle-induced upshift of the primary resonance mode manifests as a downshifted resonance field for the fundamental mode since the reinforcing effect of $H_p$ reduces the magnitude of the external field which must be applied to obtain the resonance condition. In Fig.~\ref{funiform}c  we show the consistently negative shift of the fundamental resonance for increasing particle coverages. Different particle coverages were obtained  by applying consecutively higher concentrations of particle-containing solutions to the film and measuring the FMR spectra between each application. Note that the shift is highest for intermediate particle coverages (Figs.~\ref{funiform}(f-g)) where there are large numbers of isolated particles or clusters of particles on top of the layer. In contrast, low particle coverages (Figs.~\ref{funiform}(d-e)) lead to a low reinforcing $\hp$ fields when averaged across the film (and thus low shifts). The quasi-continuous particle layers at high coverages also generate relatively small shifts. This is likely because they approximate continuous layers which, ignoring the effects of roughness, generate negligible stray fields (except at the layer boundaries).

The simulated resonance frequency shift is 34 MHz ($\equiv 0.7$ mT given an experimentally measured slope of 49.4 GHz/T for the fundamental mode, see Supporting Fig.~1). This is comparable to the maximum experimentally observed field shift in the continuous layer ($1.3\pm 0.7$ mT). Thus, we can conclude that the weak shift observed for the continuous layer is not a result of a low field sensitivity of the resonance (since it is actually quite high at almost 50 GHz$/$T), but rather due to the dominant signal coming from portions of the film subject to the relatively weak reinforcing $\hp$ fields surrounding  the particles (i.e. regions `1' and '3' in Fig.~\ref{funiform}(a)) rather  than the more intense $\hp$ fields located directly beneath the particles . 

\subsection{Patterned films: hole arrays} 

\begin{figure*}[htbp]
\centering
	\includegraphics[width=12cm]{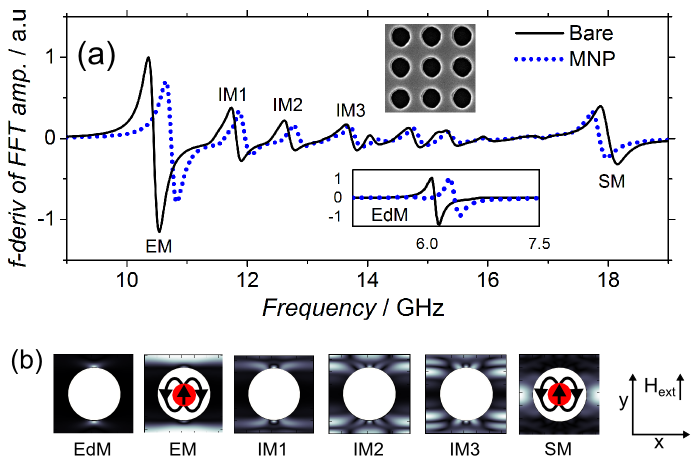}
	\caption{(a) Simulated frequency resolved FMR spectra obtained at $\hext$ = 180 mT applied in y-direction for a 300 nm MC with (dotted blue lines) and without (solid black line) a particle inside the hole. The upper inset shows an SEM image (1.3 $\mu$m x 1.3 $\mu$m ) of the MC and the lower inset shows the simulated resonance shift of the EdM in the presence of a particle. (b) Spatial concentration of dynamics for a number of resonance modes within a unit cell of the MC.  A schematic of a particle and its stray field has been added to the EM and SM profiles. }
	\label{fmodes}
\end{figure*} 

We now turn to the hole-containing nano-patterned MC where it is possible to have localized resonances in regions where $\hp$ is large. 
The main panel of Fig.~\ref{fmodes}(a) shows a simulated ferromagnetic resonance spectrum at $\hext = 180$ mT for a MC with 300 nm wide holes on a 450 nm square lattice.  The excitation spectrum is clearly much richer than the continuous layer with a number of  modes, each having different localizations within the unit cell (Fig.~\ref{fmodes}(b)). The simulated traces have been differentiated with respect to frequency for more natural comparison to  experimental FMR traces.   We will focus predominantly on the side mode (SM, shown also in Fig.~\ref{f1}(d)) and extended mode (EM) with intermediate (IM) and edge modes (EdM) discussed only briefly. While the SM is largely localized between horizontally neighboring holes, the EM occurs over extended bands running between rows of holes orthogonal to the applied field (Fig.~\ref{fmodes}(b)). IM1-3 have  similar localizations to that of the EM. Good agreement is found between the simulation and experiment for both the overall mode spectrum and the frequencies of the EM and SM (Supporting Figure 2). 

Depending on the spatial localization of each mode, the stray magnetic field from magnetized particles within the holes has a $y-$component which  can locally reinforce or oppose the $y-$oriented external field\cite{Ding2013,Metaxas2015}. This is shown schematically for the SM and EM in Fig. \ref{fmodes}(b) (one can consider the more localized modes as being nano-scale dynamic probes for the particle stray fields). Since the $y-$component of $\hp$ opposes $\hext$ where the SM is localized, the mode's frequency reduces due to a reduced net field at that location. This is  seen in Fig.~\ref{fmodes}(a) where we have also included the simulated resonance spectrum in the presence of a 150 nm wide particle at the lateral center of an anti-dot. In contrast,  the EM (and EdM) resonances  shift upwards in frequency since $\hp$ reinforces $\hext$ at the upper/lower parts of the unit cell. Note that the confinement of the modes in well-defined regions close to the particle where the stray fields are strong (i.e.~directly to the sides of the particle ($\pm x$) and directly in front of/behind it ($\pm y$)) mean that significant shifts ($\sim 0.1$ GHz) are induced for both the EM and SM (even though the sensitivities of the EM, SM and fundamental modes to static, uniform external fields are of the same order). The EdM and IMs are also subject to clear positive shifts  in the presence of a hole-localized particle, consistent with the fact that they, like the EM, have their dynamics concentrated in the upper/lower portions of the unit cell (subject to $\hp >0$).

\begin{figure*}[htbp]
\centering
	\includegraphics[width=16cm]{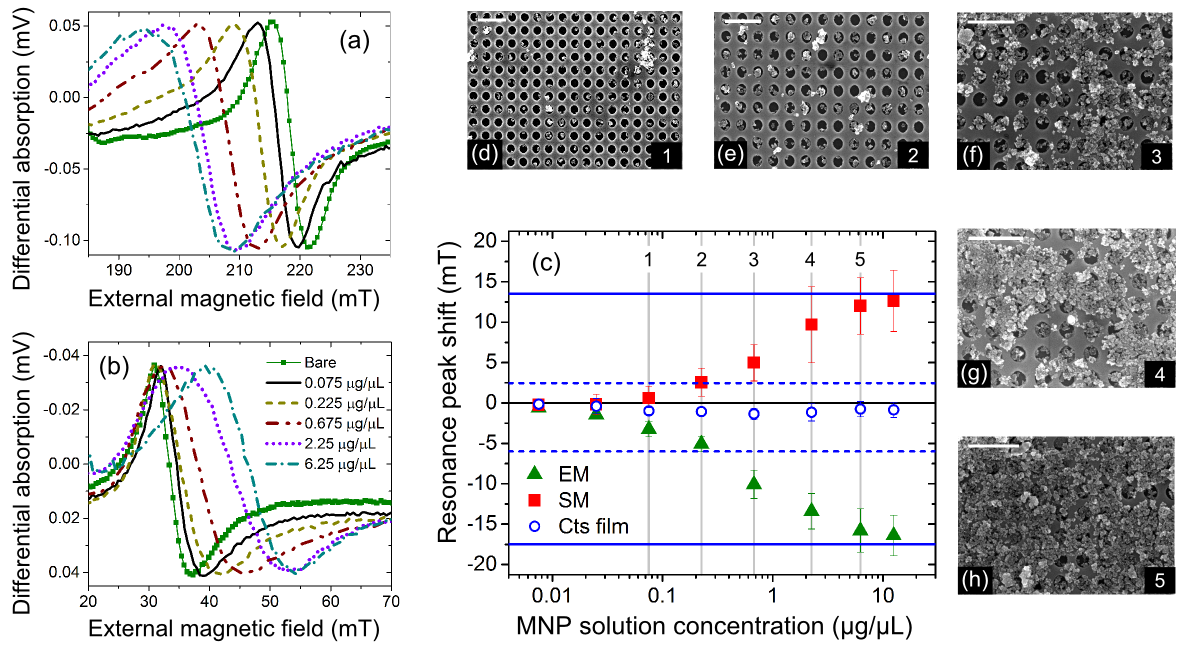}
	\caption{Experimental FMR traces (obtained at 12 GHz) showing shifts in (a) EM and (b) SM resonance lines with increase in particle concentration.(c) Experimentally obtained result for EM and SM resonant shifts for a 300 nm MC as a function of particle concentrations. Horizontal (dashed) lines show the simulated EM and SM resonant shifts for (1 particle/hole) a 150 nm thick sheet of particles on the surface. The error bars  are a measure of the spread of shifts across the measured frequency range (11.5 - 16 GHz) and also includes the uncertainties related to slight variations in sample placement ($\sim 0.5$ mT at most). The continuous film's fundamental mode shift versus particle concentration (Fig.~\ref{funiform}c) is plotted (blue open circles) for direct comparison with the patterned film. (d-h) SEM images showing clustered shaped nanoparticles inside holes and on the upper surface of the MC. White scale bars are 1 $\mu$m long. The first four data points for the EM and SM in (c) as well as the bare and 2 lowest concentration traces in (a)  have been presented previously \cite{Metaxas2015}. }
	\label{f130}
\end{figure*}

\begin{figure*}[htbp]
\centering
	\includegraphics[width=16cm]{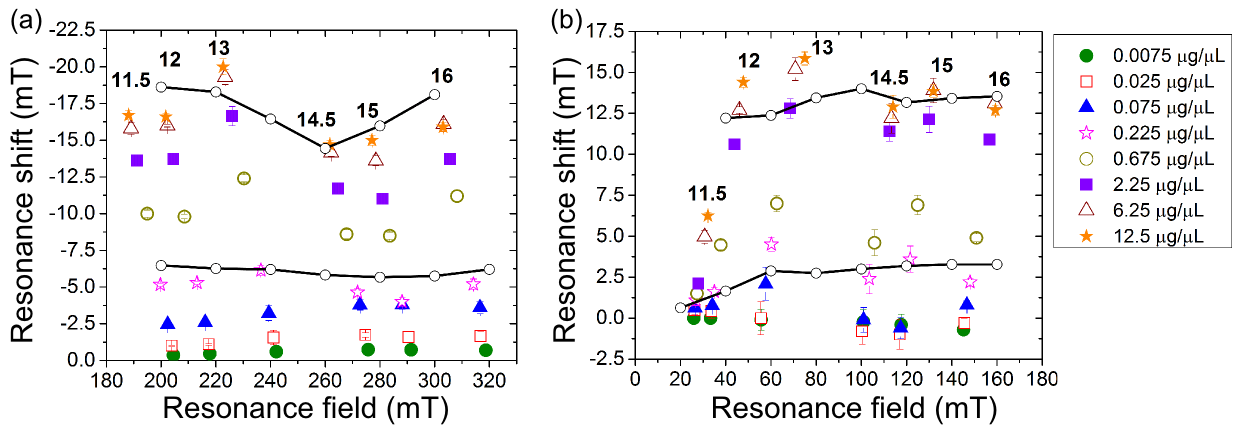}
	\caption{Resonance field shifts versus field for (a) the EM and (b) SM in the 300 nm MC. The numbers shown on the plots give the r.f.~excitation frequency in GHz. Simulated EM and SM mode shifts for the case of one particle/hole and a disk with an overlying 150 nm thick layer of particles on the crystal's surface are shown on the plots as joined points. The simulated frequency shifts obtained were converted to fields using respective mode's slopes (refer Supporting Figure~2(a)). Qualitatively good agreement is seen between the trend of simulated and experimental resonance shifts over the measured field range. }
	\label{fawesome}
\end{figure*}

Following the same protocol as for the continuous layer in the previous section, we measured the SM and EM shifts versus particle concentration in a 300 nm MC experimentally. The EM resonance field decreases since $\hp$ increases the local field (Fig.~\ref{f130}(a)). The consequence of this is that the resonance can be attained experimentally at a lower $\hext$. The resonance fields of IM1-3, which have an EM-like localization, also shift downward in experiment (Supporting Figure 4) In contrast, the SM resonance field increases because  $\hp$ locally shields the SM from $\hext$ (Fig.~\ref{f130}b). This means that a larger $\hext$ must be applied to attain the resonance condition.  As seen previously for low coverages\cite{Metaxas2015}, we observe a continuing increase of the shift magnitudes with particle coverage over a very wide range of coverages (Fig.~\ref{f130}(c)). For lower nanoparticle solution concentrations, the majority of particles are found within the holes (Figs.~\ref{f130}(d,e)) with both the percentage of filled holes and the percentage of particles lying within the holes increasing with the concentration of the particle solution (Supporting Figure 3). Saturation of the shifts commences at the penultimate concentration where a quasi-continuous layer of particles  form (Figs.~\ref{f130}(g,h)).  Notably, the maximum shifts in the MC are, as expected, significantly higher (about $10\times$) than the maximum shift observed in the continuous layer. Note that the continuous layer data from Fig.~\ref{funiform}(a) has been plotted with the MC data in Fig.~\ref{f130}(c) to enable direct comparison. Another point to note is that while the continuous layer resonance field shifts decreases when a quasi-continuous particle sheet forms, in the MC, sheet formation increases the observed shifts, an effect which is reproduced below.

We first discuss the reproduction of the observed shift for the EM and SM  (Fig.~\ref{f130}(c)) at 0.225 $\mu$g.$\mu$L$^{-1}$ where there is approximately 1 particle in each hole (e.g.~Fig.~\ref{f130}(e)). This could be done by adding a 150 nm wide particle to the simulation at the hole center and observing the resultant resonance shifts.  A good reproduction of the shifts observed at high coverages could also be obtained. There, we replaced the particle with a disk (diameter 40 nm less than the hole diameter) which was covered by a contiguous 150 nm continuous layer. This successfully approximated the effect of the well filled-holes (e.g.~Figs.~\ref{f130}(g,h)) covered by a semi-complete particle sheet, as observed at high particle solution concentrations.  The agreement between simulation and experiment for the EM and SM can be seen in Fig.~\ref{f130}(c) which shows the field-averaged simulated shifts for the EM and SM for one particle per hole (horizontal dashed line) and full particle coverage (solid dashed line). There is excellent  agreement between these values and the shifts experimentally obtained respectively at the fourth  and final coverages. 

Experimental and simulated shifts obtained at different fields are shown in Fig.~\ref{fawesome}. Experimentally, different resonance fields are obtained by changing the r.f. frequency and it is the data in this figure which has been averaged to obtain the points in Fig.~\ref{f130}(c). Here one can again see explicitly that the resonance fields decrease with coverage at each frequency for the EM (leading to leftward slanting data in Fig.~\ref{fawesome}(a)) but increase for the SM (leading to rightward slanting data in Fig.~\ref{fawesome}(b)). The simulated shifts obtained explicitly for different $\hext$ values, taking into account the field-dependent moment of the particles\cite{Metaxas2015} could notably also reproduce the experimentally observed field-dependencies of the resonance field shifts at the fourth and final coverages (Fig.~\ref{fawesome}). The most intuitive of these dependencies is a reduced shift at very low fields for the SM mode where the field-dependent particle moment (and thus resultant stray field) is weak (Fig.~\ref{fawesome}(b)). In general though, increasing the field does not strongly change the size of the shifts meaning that the external field can be incnreased to maximize the particle moments without compromising the frequency shift.

\begin{figure*}[htbp]
\centering
	\includegraphics[width=16cm]{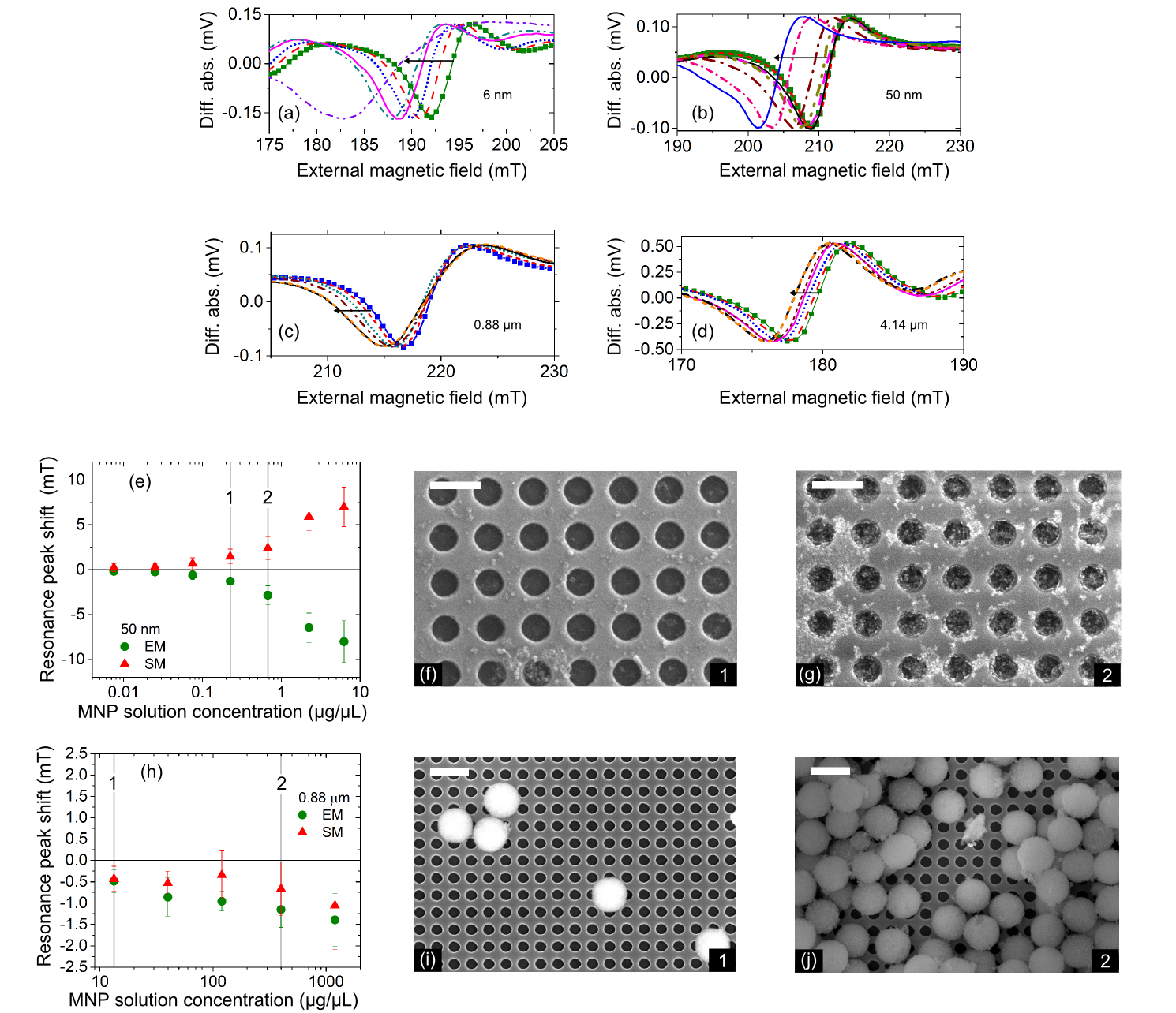}
	\caption{Experimentally obtained EM FMR traces (excited at 12 GHz) showing decrease in resonance field with increase in concentration of (a) 6 nm nanoparticles, (b) 50 nm nanoparticles, (c) 0.88 $\mu$m magnetic beads and (d) 4 $\mu$m magnetic beads. (e) EM and SM resonant shifts as a function of applied 50 nm particle concentrations. SEM images showing the distribution of 50 nm particles for (f) lower and (g) higher concentrations. (h) EM and SM resonant shifts as a function of applied 0.88 $\mu$m bead concentrations. SEM images showing the distribution of 0.88 $\mu$m  wide magnetic beads for (i) lower and (j) higher concentrations. The error bars in (e,h) are a measure of the spread of shifts across the measured frequency range (11.5 - 16 GHz). White scale bars are 1 $\mu$m long.}
	\label{fMNPsize}
\end{figure*} 

We finally note that particle sensing was also demonstrated with a second MC having smaller holes (240 nm) but the same lattice pitch.  Analogous effects were observed except that the shifts were reduced. The shifts could albeit be reproduced by assuming a filling proportional to the hole size (Supporting Figure 5). This is consistent with the reduced shifts being due to a reduced capability of filling rather than an intrinsically lower field sensitivity of the modes. Indeed the intrinsic field sensitivities are similar to the MC with 300 nm holes (Supporting Figure 2).

\subsection{The role of particle size and localization}

In Figs.~\ref{fMNPsize}(a-d), we show shifts in the EM resonance field induced by particles with (average or supplier-stated) diameters of 6 nm, 50 nm, 0.88 $\mu$m and 4 $\mu$m. Clear shifts are observed in all cases, demonstrating the compatibility of this sensing method with a very large range of particle sizes.  As explained below however, the polarities of the observed shifts  depends on the size of the particles relative to the size of the holes. Before concentrating on the shift polarities, we briefly note that as  seen in Figs.~\ref{f130}(a,b) and \ref{fMNPsize}(a-d),  the dominant effect of both nanoparticles and beads is typically a resonance shift, albeit accompanied by a relatively weak degree of linewidth broadening (discussed below). Exceptions to this are the 0.88 $\mu$m particles and the highest coverage scenario for the 6 nm particles where broadening and shifts are more comparable.

The same opposing SM and EM shift directions seen in Fig.~\ref{f130}(c) for the 130 nm particles were  observed for two other  particle types, the common aspect of these particles being that they could enter the MC's holes thanks to their small size: 6 nm wide particles (Fig.~\ref{fMNPsize}(a) and Supporting Figure 6) and 50 nm wide particles (Figs.~\ref{fMNPsize}(b,e-g)). In both cases, the overall trend is again a stronger shift with increasing particle coverage. This is seen clearly in Figs.~\ref{fMNPsize}(a,b) showing consecutively larger shifts of the EM resonance for both particle sizes.  Note that a clear SM shift was seen for the 6 nm particles only at the highest concentration (Supporting Figure 6). 

\begin{figure}[htbp]
\centering
	\includegraphics[width=7cm]{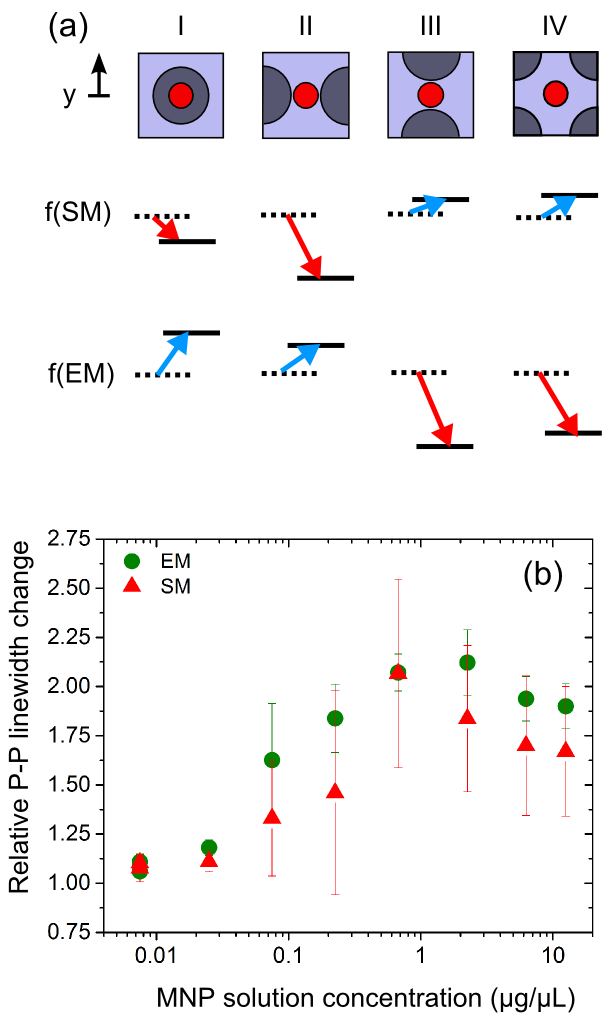}
	\caption{ (a) Simulated SM and EM frequency shifts measured with respect to the bare frequency (shown as dotted lines) for a 150 nm particle located within the MC hole (I) and at various positions on the MC surface (II: above the SM; III, IV: above the EM). Significant reductions in a mode's frequency are observed when a particle is above the region of the MC containing that mode since $\hp$ strongly opposes $\hext$ (position `2' in Fig.~\ref{funiform}(a)). Weaker increases in the other mode are observed since $\hp$ will weakly reinforce $\hext$ (positions `1' and `3' in Fig.~\ref{funiform}(a)). Although the induced shifts can be large, the observation of such a shift will be  highly dependent on having this particular particle placement.  Note that, vertical distances in the schematic are proportional to the simulated frequency shifts. (b) Experimentally observed changes in EM and SM resonance linewidths as a function of 130 nm particle concentrations applied to 300 nm MC. The error bars  are a measure of the spread of linewidth broadening across the measured frequency range, 11.5 - 16 GHz. Data for the lowest four concentrations are taken from \cite{Metaxas2015}.}
	\label{f5}
\end{figure}

As already highlighted for the 130 nm particles (and seen clearly in Fig.~\ref{fMNPsize}(g) for the 50 nm particles) there can also be significant numbers of isolated (groups of) small particles on the MC surface which can of course also generate resonance shifts. Indeed, these shifts can be comparable in magnitude to those observed for hole-localized particles, especially when the particles are directly above the position where a mode's dynamics are concentrated. In contrast to the fundamental mode in the continuous film, due to geometrical confinement, the modes in the MC are inherently localized and indeed can be localized directly beneath a particle. In such a situation,  the whole region containing the mode will be subject to the strong $\hp$ below the particle which opposes $\hext$   (position `2' in Fig.~\ref{funiform}(a)),  thus decreasing the resonance frequency. This has been shown explicitly via simulation for three different surface positions in Fig.~\ref{f5}(a) where simulated shifts of the EM and SM mode due to surface-located particles are compared to the shifts induced by a particle at the center of the hole. For example, when a particle is above the SM mode, shown as case I in Fig.~\ref{f5}(a), its frequency is strongly downshifted due to $\hp$ locally opposing $\hext$ (the SM is localized in region `2' in Fig.~\ref{funiform}(a)). In this case the EM frequency will be slightly upshifted (the EM will be localized in regions `1' and '3' in Fig.~\ref{funiform}(a) where $\hp$ is weaker but reinforces $\hext$). Similarly, when the particle is lying above the EM, the EM frequency is strongly downshifted whereas the SM frequency is slightly upshifted (cases III and IV in Fig.~\ref{f5}(a)).  The critical point to remember here though is that the placement of small particles on the MC's upper surface is quite random within the unit cell. As a result, and as observed in experiment,  the observed shifts will be dominated by hole-localized particles due to their common, preferential in-hole localization. As already shown though, the effect of a quasi-continuous layer of particles is such that it reinforces the effect of  hole-localized particles. 

In Fig.~\ref{f5}(b), we show the linewidth broadening observed for the SM and EM due to application of the 130 nm particles onto the 300 nm MC. Notably, broadening is highest (increase of $\sim \times 2$) around the 6th coverage (Fig.~\ref{f130}(g)) where there is a large number  of isolated particles (or groups of particles) on the MC's upper surface. These mid-to-high coverages lead to relatively high numbers of isolated particles on the surface which can locally modify the underlying modes, as discussed above. Given the position-dependence of the shifts (Fig.~\ref{f5}(a)), linewidth broadening is not  unexpected since surface-located particles can both increase and decrease the resonant fields, depending on their location (Fig.~\ref{f5}(a)). The formation of a quasi-continuous layer of particles at higher coverages however, would be expected to reduce the strong localized fields  generated underneath the particles. Indeed, at these higher coverages, there is a slight reduction of the linewidth  (Fig.~\ref{f5}(b)) consistent with a reduced contribution from isolated surface-located particles. We also note that distributions in the particle sizes (as can be identified in Figs.~\ref{f130}(d-h)) will also contribute to linewidth broadening since smaller (larger) particles will generate smaller (larger) shifts (Supporting Figure 7). This broadening will be in a common direction however since both smaller and larger hole-localized particles generate shifts of the  polarity shift. 

We now turn to particle diameters which are larger than the hole diameter, 0.88 $\mu$m (Fig.~\ref{fMNPsize}(c,h-j)) and 4 $\mu$m (Fig.~\ref{fMNPsize}(d) and Supporting Figure 6), we see that both the SM and EM modes are characterized by negative polarity field shifts. This mirrors the behavior observed for the fundamental mode in the continuous layer, suggesting that, for large particles which are not predominantly hole-localized, the dominant shift, as for the continuous layer, comes from in-plane field components, here generated by large microbeads quasi-randomly distributed  on the MC surface. Indeed, despite some centering of the 0.88 $\mu$m particles on top of holes at low concentrations (two bottom right particles in Fig.~\ref{fMNPsize}(i)), they are indeed typically randomly distributed across the MC. 
Unlike the case of the small particles however, there are no hole-localized particles which can act in unison to induce different polarity shifts on the EM and SM. As a result, for purely surface-localized beads, both the EM and SM shift in the same direction.

\section{Conclusions}

Magnetic particle detection is critical for magnetic biosensing techniques which have envisioned use in point of care medical diagnostics. This detection is  often carried out using conventional magnetoresistive sensors. Looking towards the development of alternative frequency-based sensing methods however,  we have demonstrated here the use of nano-confined ferromagnetic resonance modes for magnetic particle sensing. This has been carried out for  a range of magnetic particles with sizes from 6 nm to 4 $\mu$m. The stray fields of the particles act directly on the precessing moments generating significant shifts in the resonance fields and frequencies.  As such, the observed resonance shifts in our hole-based patterned ferromagnet can  be maintained over large ranges of fields, enabling resonance-based nanoparticle detection in strong fields where particle moments can be maximized.
Furthermore, the dominant effect is typically a resonance shift rather than a broadening of the resonance linewidth, a result which is encouraging for the implementation of a sensing method based on detecting changes to resonance frequencies. 

 In this work, we also  identified different characteristics in the resonance shifts for small and large particles.
For small particles, their preferential capture in our system's nano-holes  means that the majority of them act on the resonance modes in unison, generating clear, mode-dependent resonance shifts and mode-dependent shift polarities which persist up to extremely high particle coverages. Shifts at intermediate and large coverages of small particles can be well reproduced via simulation. In contrast to the case of the small particles however, the effect of microbeads is analogous to that seen in continuous films with the two major resonance modes both shifting weakly in the same direction. This is due to a lack of particle localization with respect to the spatially periodic nano-scale regions where the modes are localized. Although sub-optimal, detection was nevertheless achieved in such geometries, albeit with low resonance shifts which were similar to those seen in continuous films. Note that the ability to detect particles is determined not only by how sensitive the frequency is to changes in magnetic field but by the ability to generate clear modifications to these frequencies due to favorable particle positioning. A proper choice of particle size and reliable positioning  of particles  close to well localized modes is thus a critical factor in resonance based sensing.

\section{Experimental section}

The nanopatterned films consist of square arrays of circular holes (`anti-dots') with an array pitch of 450 nm and hole diameters of  240 nm or 300 nm in a 30 nm thick Permalloy (\nife) continuous film with a 8 or 10 nm gold (Au) capping layer. The large area ($4\times 4$ mm$^2$) MCs were fabricated on silicon (Si) substrates using deep ultraviolet lithography, e-beam deposition and liftoff.\cite{Singh2004} 

FMR  was measured using broadband microwave stripline-based FMR spectroscopy, a technique where resonant modes are excited in magnetic materials using a radio-frequency (r.f.) field, here generated by an r.f.~signal passing through an underlying stripline. An absorption of microwave power by the magnetic sample is measured when the frequency of the r.f.~field matches that of a FMR mode. Here we used field-modulated FMR spectroscopy wherein a lock-in amplifier measures the external field-($\hext$-)derivative of the FMR response at a constant frequency while sweeping $\hext$. An interferometric receiver was used  to maximize signal amplitudes.\cite{Ivanov2014} This is particularly important in our measurement since the sample is separated from the stripline by a microscope coverslip (rather than sitting directly on the stripline). The coverslip ensures that particles do not rub off onto the micro-stripline but reduces the signal amplitude. $\hext$ was typically swept from 0-350 mT. We note that an overall decrease in the differential FMR signal was seen when adding increasing concentrations of particles. To enable comparison of traces, the traces were vertically scaled. A small vertical offset (typically on the order of a few tens of $\mu$V at most) at times also had to be corrected for.

Particle detection was carried out for a range of particles: (i) $6\pm 1$ nm wide iron-oxide nanoparticles  (ii) nanomag-D(-spio) cluster-shaped particles from Micromod Partikeltechnologie GmbH with stated diameters of 50 nm and 130 nm; (iii) 880 nm wide magnetic beads from Bangs Laboratories consisting of iron-oxide nanoparticles within a polymer matrix; and (iv) 4 $\mu$m  beads from SpheroTech consisting of a polystyrene core coated with a mixture of polystyrene and magnetic nanoparticles.  Particles were applied in solution to the films' upper surfaces in ambient laboratory conditions in the absence of an external magnetic field ($\hext=0$). The solutions were then allowed to dry before re-measurement and imaging.  Concentrations of commercial particle solutions (all aqueous) were determined from the manufacturer's specifications after dilution in purified water. The 6 nm iron-oxide nanoparticles were synthesized under standard Schlenk conditions using the methodology developed by Sun \textit{et al.}.\cite{Sun2004} Briefly, Fe(acac)$^3$ (0.7 g, 2 mmol) and 1,2-hexadecanediol (2.5 g, 10 mmol) were dissolved in benzyl ether (20 mL) containing oleylamine (6 mmol) and oleic acid (6 mmol). The resulting mixture was heated at 200 $^{\circ}$C for 2 hours, increased to 260 $^{\circ}$C, held for 1 hour then cooled to room temperature. The particles were precipitated by the addition of ethanol (40 mL), centrifuged (5000 rpm, 10 mins) and redispersed in 1,2-dichlorobenzene to the required concentration. The particles have been fully characterized by a range of routine techniques. The results are contained in Supporting Figure 8.

Micromagnetic simulations were run for a single unit cell of the MC (or an equivalently sized region of conintuous film), employing periodic boundary conditions and a tiled macro-geometry\cite{Fangohr2009} ($33\times 33$ unit cells). For nanoparticles within the holes of the MC, the lower surface of the particle was always located  at  the lower surface of the MC (i.e.~resting on the underlying Si wafer). For particles on top of the MC surface or above the layer, the lower surface of the particle was set at 10 nm above the upper surface of the magnetic media (i.e.~sitting on the upper surface of the non-magnetic capping layer). The following parameters were used for the MC: damping $\alpha$ = 0.008, nil intrinsic anisotropy, gyromagnetic ratio 2$\pi\gamma$ = $1.85 \times 10^{11}$ rad/T.s, saturation magnetization $M_s$ = $8 \times 10^5$ A/m and exchange stiffness $A_{ex}$ = 13 pJ/m.  The parameters for the particle were: damping $\alpha$ = 0.05, gyromagnetic ratio $\gamma$ as in the MC and $M_s$ taken from previously presented nanoparticle magnetization data  \cite{Metaxas2015}. Micromagnetic results were obtained from time domain simulations run using MuMax3 \cite{Vansteenkiste2014}. For these simulations, the system was initialized with a uniform magnetization in the (1,1) direction and allowed to relax in the presence of an external field using MuMax3's internal relaxation routine. Post-relaxation, an excitation sinc pulse of 0.5mT (cut-off freq 30 Ghz, 300 ps offset) was applied along the $x$-axis\cite{Venkat2013} at a given $\hext$ (applied along the $y$-axis). The precessional ringdown data was then Fourier analyzed. All resultant mode profile visualizations were determined by extracting the spatially resolved $m_x$ Fourier amplitudes at each identified resonant frequency across the simulation region. Obtained results (e.g.~Fig.~\ref{fmodes}(b)) are shown as intensity plots with the brightest regions corresponding to the highest Fourier amplitude for $m_x$ at that frequency. matplotlib\cite{Hunter2007} was used for visualization of simulation data. As done previously\cite{Metaxas2015}, the eigensolver (see e.g.~\cite{Metaxas2016}) in the FinMag micromagnetic simulation package (based on Nmag\cite{Fischbacher2007}) was also used for certain test cases with good levels of agreement.

\section{Acknowledgments}
This research was supported by the Australian Research Council's Discovery Early Career Researcher Award scheme (DE120100155) and Discovery Projects scheme (DP110103980), the United States Air Force (Asian Office of Aerospace Research and Development, AOARD) an EPSRC DTC grant (EP/G03690X/1), the University of Western Australia's (UWA)  RDA, RCA,  ECRFS, SIRF, UPAIS, Re-Entry Fellowship, Teaching Relief and Vacation Scholarship schemes and by resources provided by the Pawsey Supercomputing
   Centre with funding from the Australian Government and the Government of
   Western Australia. A.O.A. was supported by the National Research Foundation, Prime Minister's Office, Singapore under its Competitive Research Programme (CRP Award No. NRF-CRP 10-2012-03). The authors thank C.~Lueng, M.~Albert, W.~Wang, A.~Suvorova,  A.~Dodd, C.~Yang, D.~Schibeci, A.~Chew and R.C.~Bording   for their assistance. The authors  acknowledge access to the UWA's Biomagnetics Wet Laboratory and Magnetic Characterisation Facility as well as the facilities, and the scientific and technical assistance of the Australian Microscopy \& Microanalysis Research Facility at the Centre for Microscopy, Characterisation \& Analysis, The University of Western Australia, a facility funded by the University, State and Commonwealth Governments.


\newpage

\section*{Supporting figure 1}
 
\begin{figure*}[htbp]
	\centering
	\includegraphics[width=7cm]{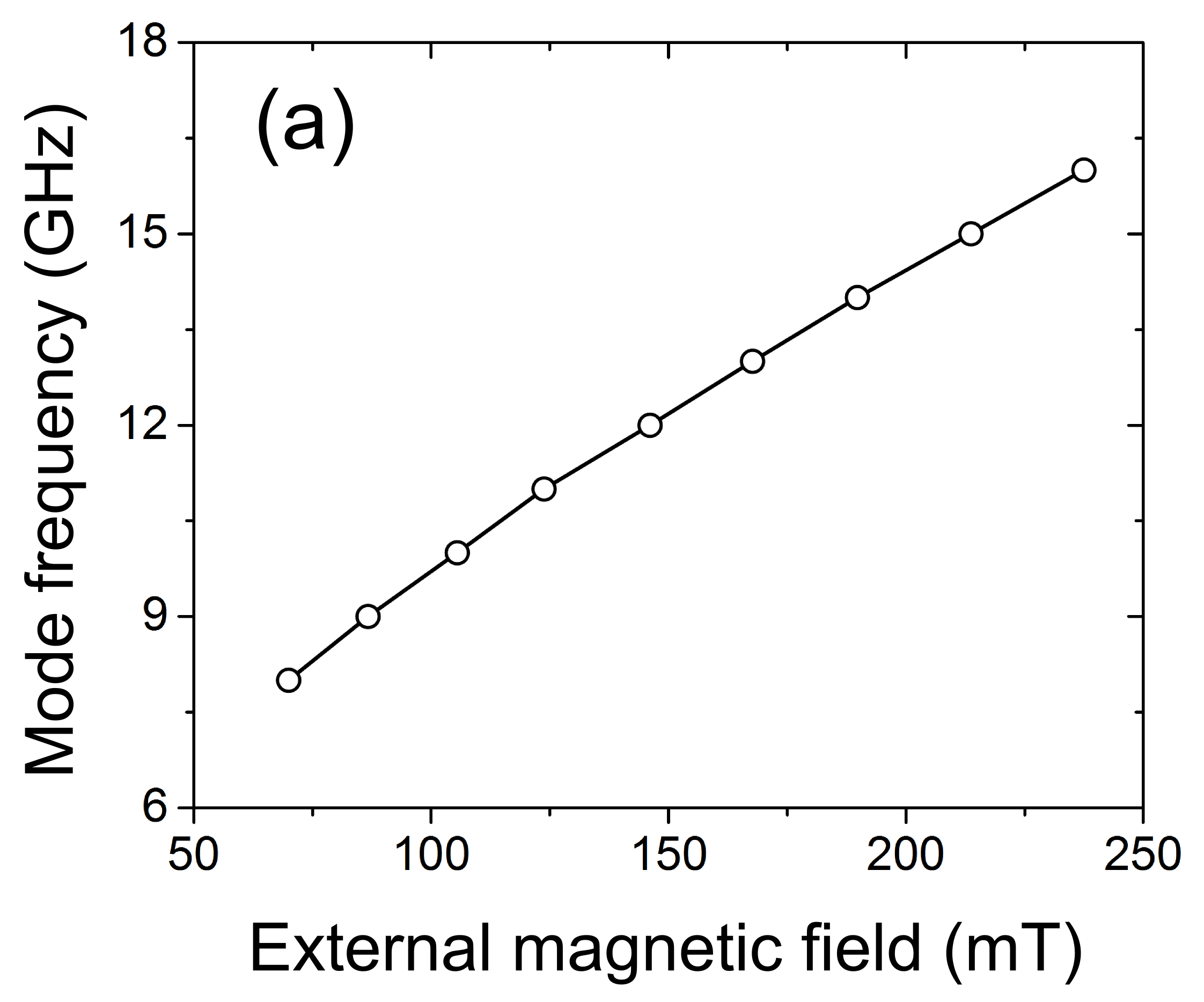}
	\label{supp1}
\end{figure*}
 
\noindent Fundamental mode resonance frequencies of the bare continuous magnetic film  plotted versus $\hext$. An average slope of 49.4 GHz/T was used to convert between particle-induced field shifts and frequency shifts.

\newpage

\section*{Supporting figure 2}

\begin{figure*}[htbp]
	\centering
	\includegraphics[width=14cm]{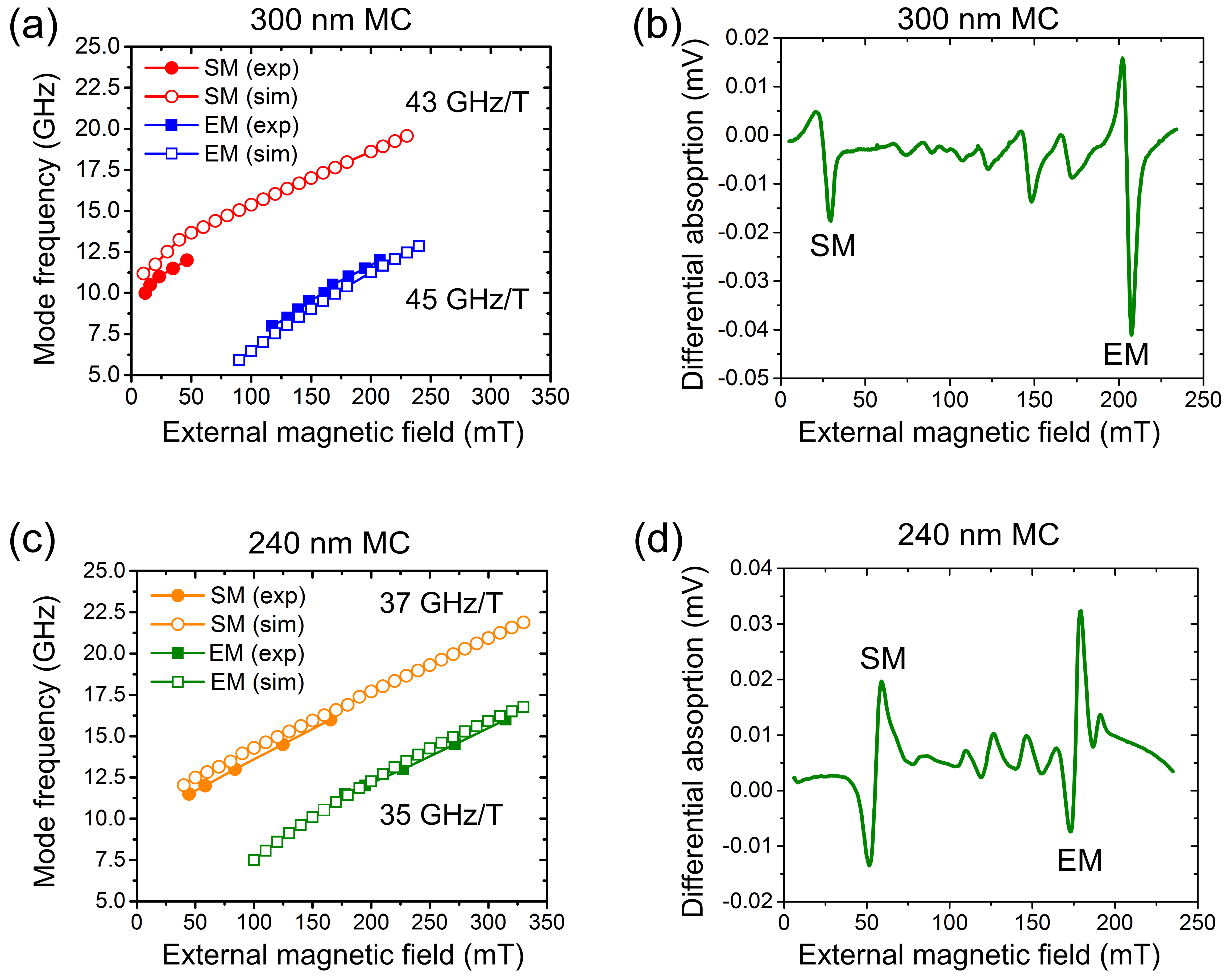}
	\label{supp2}
\end{figure*}

\noindent (a,c) show  comparisons between the simulated and experimental mode structures (EM and SM) for both studied MC geometries (240 nm and 300 nm holes on a 450 nm array pitch). The slopes of the data were used to convert between particle-induced field shifts and frequency shifts. (b,d) shows experimentally obtained FMR traces of bare MCs at a fixed r.f.~excitation frequency of 12 GHz. Intermediate modes are clearly resolved in the plots, lying between the SM and EM.  

\newpage

\section*{Supporting figure 3}

\begin{figure*}[htbp]
	\centering
	\includegraphics[width=14cm]{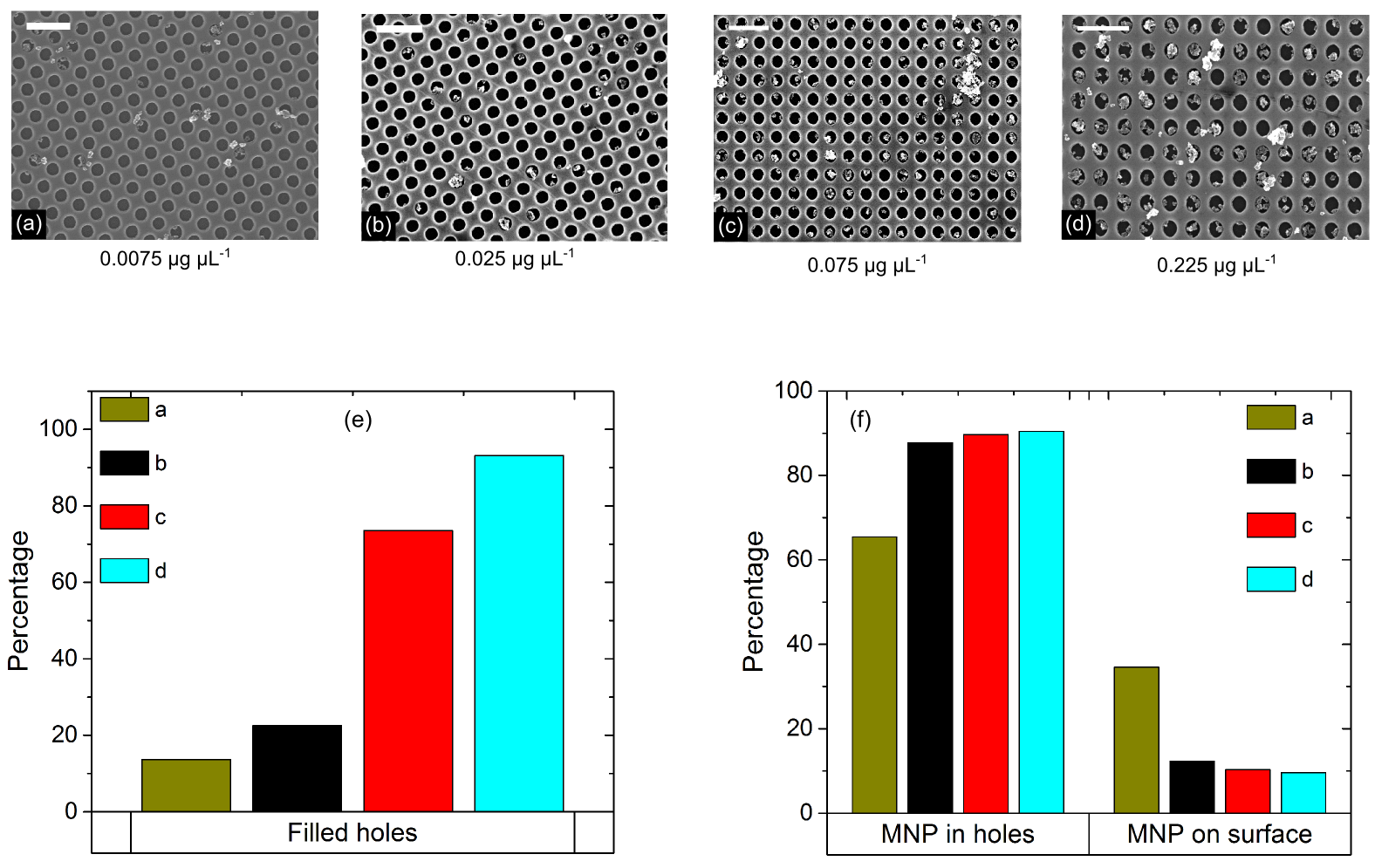}
	\label{supp3}
\end{figure*}

\noindent (a-d) SEM images  showing the four lowest concentrations of 130 nm particles on the 300 MC. These images were used to extract statistics regarding the number of particles inside holes and  on the MC surface. (e) shows the ratio of filled holes versus particle solution concentrations. (f) shows that increasing the particle solution concentration also increases the proportion of hole-localized particles.

\newpage

\section*{Supporting figure 4}
\begin{figure*}[htbp]
	\centering
	\includegraphics[width=6cm]{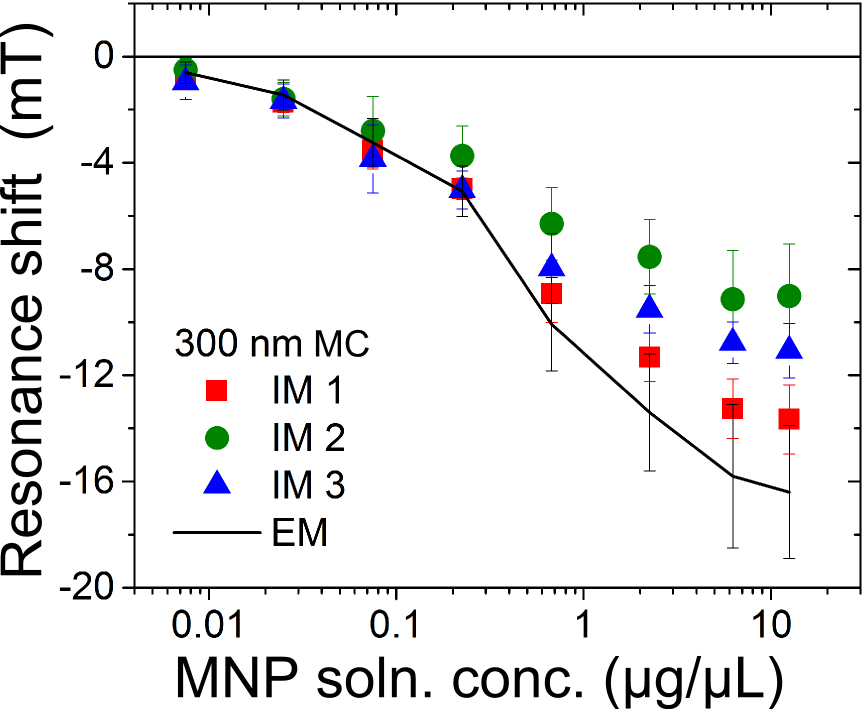}
	\label{supp4}
\end{figure*}

\noindent Experimentally obtained resonance shifts for intermediate modes (IM1-3) as a function of  particle solution concentration (130 nm particles). The solid line shows the EM shifts for comparison.
\newpage

\newpage
\section*{Supporting figure 5}
\begin{figure*}[htbp]
	\centering
	\includegraphics[width=14cm]{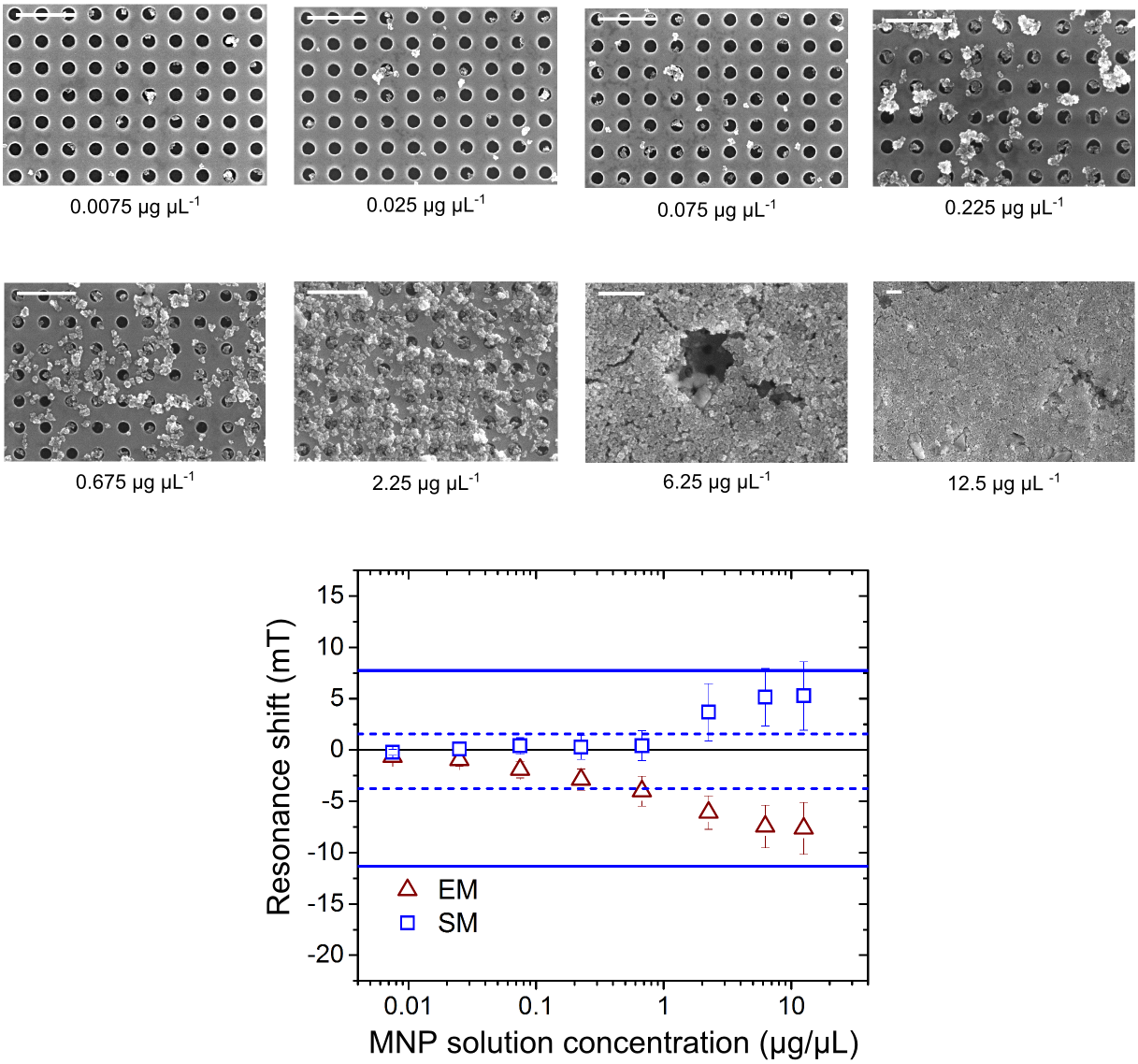}
	\label{fpat}
\end{figure*}

\noindent Upper: SEM images of 130 nm particles on a 240 nm MC. Particle concentrations are given below the images. The white scale bar represents 1 $\mu$m. Lower: Experimentally obtained result for EM and SM resonant shifts for a 240 nm MC as a function of particle concentration. Horizontal dashed and solid lines resepctively show the average simulated EM and SM resonant shifts for 1 particle/hole and for a disk plus a 150 nm thick sheet of particles on the surface (as described in the manuscript). The error bars  are a measure of the spread of experimentally obtained shifts across this frequency range and also includes the uncertainties related to slight variations in sample placement.
\newpage

\section*{Supporting figure 6}
\begin{figure*}[htbp]
	\centering
	\includegraphics[width=14cm]{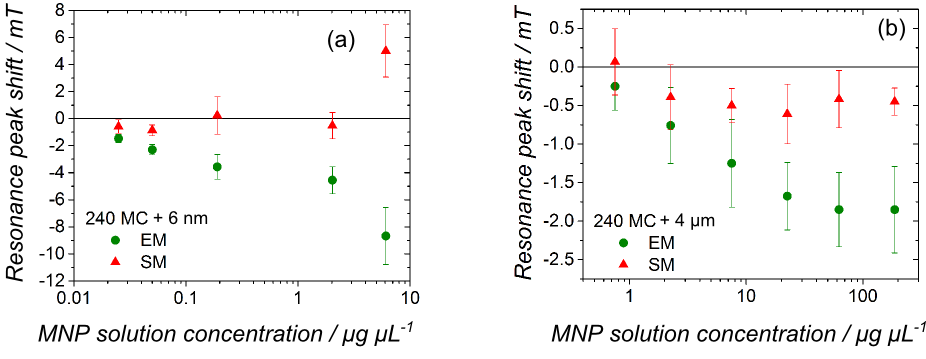}
	\label{supp7}
\end{figure*}

\noindent Experimentally obtained  EM and SM resonant shifts for a 240 nm MC versus solution concentration for (a) 6 nm particles and (b) 4 $\mu$m wide magnetic beads. 

\newpage

\section*{Supporting figure 7}
\begin{figure*}[htbp]
	\centering
	\includegraphics[width=10cm]{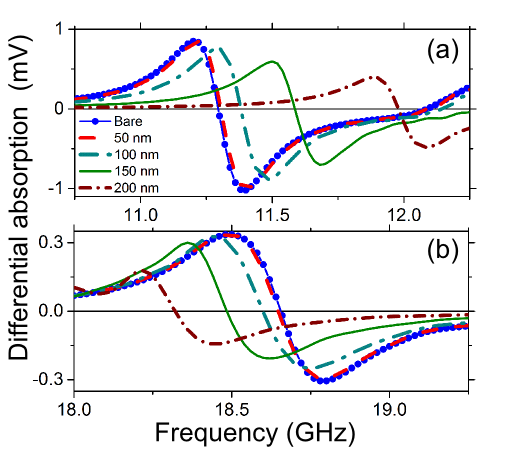}
	\label{supp9}
\end{figure*}

\noindent Simulated (a) EM and (b) SM resonance shifts in a 300 nm MC as a function of particle size for a hole-located particle laterally centered within the hole.
\newpage

\section*{Supporting figure 8}
\begin{figure*}[htbp]
	\centering
	\includegraphics[width=14cm]{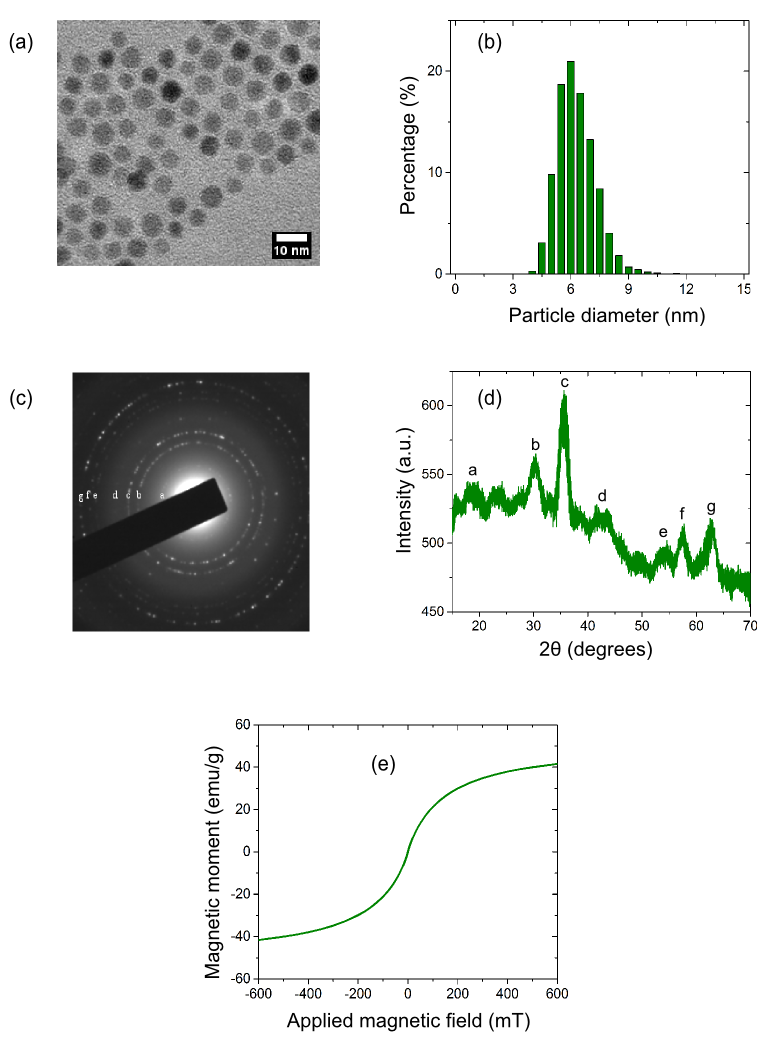}
	\label{supp10}
\end{figure*}

\noindent (a) Bright field TEM image of as-synthesized iron-oxide particles. (b) Particle size distribution  (obtained from 7096 particles). (c) Selected area electron diffraction pattern (SAED). (d) X-ray powder diffraction (XRD) spectrum. Labels correspond to the following indices: a (111); b (220); c (311); d (400); e (422); f (511) and g (440). (e) Magnetic moment of freeze-dried 6 nm particles m versus applied magnetic field.

\end{document}